\documentclass[reprint,
superscriptaddress,showkeys,
 amsmath,amssymb,
]{revtex4-2}

\usepackage{graphicx}
\usepackage{dcolumn}
\usepackage{bm}

\begin{document}

\title{High-quality level-crossing resonances under counterpropagating\\circularly polarized light waves for applications in atomic magnetometry}

\author{D.V. Brazhnikov}
\email{brazhnikov@laser.nsc.ru}

\affiliation{Institute of Laser Physics SB RAS, 15B Lavrentyev Avenue, Novosibirsk 630090, Russia}

\affiliation{Novosibirsk State University, 1 Pirogov Street, Novosibirsk 630090, Russia}

\author{V.I. Vishnyakov}%
 
\affiliation{Institute of Laser Physics SB RAS, 15B Lavrentyev Avenue, Novosibirsk 630090, Russia}

\author{S.M. Ignatovich}%
 
\affiliation{Institute of Laser Physics SB RAS, 15B Lavrentyev Avenue, Novosibirsk 630090, Russia}

\author{I.S. Mesenzova}%
 
\affiliation{Institute of Laser Physics SB RAS, 15B Lavrentyev Avenue, Novosibirsk 630090, Russia}

\author{C. Andreeva}%
 
\affiliation{Institute of Electronics BAS, 72 Tsarigradsko Chaussee, Sofia 1784, Bulgaria}

\affiliation{Faculty of Physics, Sofia University "St. Kliment Ohridski", 5 James Bourchier Boulevard, Sofia 1164, Bulgaria}

\author{A.N. Goncharov}%
 
\affiliation{Institute of Laser Physics SB RAS, 15B Lavrentyev Avenue, Novosibirsk 630090, Russia}

\affiliation{Novosibirsk State University, 1 Pirogov Street, Novosibirsk 630090, Russia}

\affiliation{Novosibirsk State Technical University, 20 Karl Marks Avenue, Novosibirsk 630073, Russia}

\date{\today}

\begin{abstract}
Level-crossing (LC) resonances in a buffer-gas-filled cesium vapor cell are studied under counterpropagating pump and probe light waves with opposite circular polarizations. The waves excite the D$_1$-line ground-state level $F_g$$=\,$$4$, while a transverse magnetic field ($\textbf{B}_x$$\perp$$\textbf{k}$) is scanned around zero to observe the resonance of electromagnetically induced absorption (EIA). It is shown that adding the pump light wave significantly improves the properties of the resonances in comparison with the commonly used scheme with a single light wave. As far as a small vapor cell ($\approx0.1$~cm$^3$) at relatively low temperature ($\approx45$$-$$60$~$^\circ$C) is utilized, the results have good prospects for developing a low-power miniaturized atomic magnetometer.
\end{abstract}

\maketitle


\section{\label{sec:Intro}Introduction}

Starting from the experiments of Wilhelm Hanle and his contemporaries, the level-crossing (LC) phenomenon in resonant atomic vapors has played a crucial role in the development of quantum mechanics \cite{Hanle1924,Breit}. The zero-field LC resonances are nowadays often studied as a change in the light wave intensity transmitted through a resonant medium when an ambient magnetic field is scanned. The level crossing in an atomic ground state -- the ground-state Hanle effect (GSHE) -- can provide much narrower magneto-optical resonances than the excited-state Hanle effect, especially if a buffer gas or an antirelaxation coating of the vapor cell walls are used \cite{Alexandrov,DupontRoc}. Such resonances were immediately proposed for measuring weak magnetic fields.

To date, there is a large variety of laser spectroscopy methods that can be applied for magnetic field measurements \cite{BudkerReview}. The GSHE-based technique does not usually claim a record sensitivity as, for instance, some types of laboratory atomic magnetometers (AMs) based on nonlinear magneto-optical rotation \cite{Kominis2003}, nevertheless, it remains a simple and robust technique, reaching a subpicotesla sensitivity. The method can provide scalar \cite{Alexandrov,DupontRoc,ShahMEMS,Osborne2018}, two-axis \cite{Papoyan1,Grewal2D,LeGal} or full three-axis (vector) measurements \cite{LeGal,Alipieva,Behera}, and the sensor can be miniaturized to a great extent \cite{ShahMEMS,Osborne2018}. Besides, the cross-talk effects occurring in a multi-channel mode of operation are relatively small in the Hanle magnetometers compared to some other types of AMs where an rf field is used. Therefore, the Hanle magnetometers are attractive for a wide range of applications, including biology and medicine \cite{Osborne2018}. The simplest GSHE-based magnetic field sensor usually utilizes a single circularly polarized light wave. The wave produces a macroscopic vector magnetization in the atomic ground state, following the $\sigma$ optical transitions and the optical pumping process \cite{Happer}. To observe the resonance, a transverse magnetic field (${\bf B}_x$$\perp$${\bf k}$ with ${\bf k}$ being the wave vector) is scanned around zero \cite{Alexandrov,Papoyan1,ShahMEMS}. In this configuration, the so-called electromagnetically induced transparency (EIT) is usually observed.

Here we describe a possibility how this standard configuration can be modified to obtain a significant improvement in the LC resonance parameters. We focus on a low temperature regime ($\leq60$~$^\circ$C) and on the use of a small vapor cell with inner volume of $\approx\, 0.1$~cm$^3$. Such requirements can be important for various applications where heat release, power consumption and compactness of a sensor play a crucial role (i.e., see the discussions in \cite{Sheng2017}). We propose to use a cesium vapor rather than rubidium, potassium or sodium ones. This choice is dictated by the facts that, on the one hand, Cs has the highest saturated vapor pressure and, on the other hand, it has the strongest hyperfine splitting (hfs) in the ground state ($\approx\,9.2$~GHz). The former means that the vapor cell can operate at a relatively low temperature (for instance, K sensors usually operate at temperatures higher than 100~$^\circ$C). The large ground-state hfs of Cs means that the atomic quantum system, during its interaction with the light field, remains open even at a relatively high buffer gas pressure. In other words, owing to the spontaneous decay process, atoms can be accumulated on the non-resonant ground-state level $F_g$$=$$3$, while the level $F_g$$=$$4$ is being excited by the light field. For a long time, many authors considered the latter as a parasitic effect. Indeed, in the standard Hanle configurations, both with circularly or linearly polarized light waves, the openness leads to serious degradation of both the EIT and the EIA properties \cite{Openness1,Openness2,Openness3}. In \cite{Gozzini2009}, the authors proposed to overcome this problem by using potassium vapors, because the K ground-state hfs equals $\approx0.46$~GHz, which is smaller than the Doppler width of $\approx0.77$~GHz at $300$~K. Consequently, the quantum system in this case is not open, i.e. both ground-state levels efficiently interact with the light field. However, potassium vapors require either a high temperature or an extended length of the cell to obtain a desirable degree of light field absorption. A similar reasoning can be applied to the sodium vapor.

In \cite{LPL2014}, we proposed an idea of adding a second (pump) light wave to improve the parameters of the GSHE resonances. The idea was successfully verified experimentally with Rb \cite{LPL2018} and Cs \cite{JPhysB2019} buffer-gas vapor cells of cm dimension. Counterpropagating waves with orthogonal linear polarizations were used (\textit{lin}$\perp$\textit{lin} configuration), leading to observation of an EIA-type LC resonance in the probe wave transmission. A unique feature of the \textit{lin}$\perp$\textit{lin} configuration is that the openness plays an essentially positive role for achieving high contrast of the resonances. In particular, in the case of Cs, a resonance contrast as high as $75$~\% was achieved when the waves were in resonance with the optical transition $F_{\rm g}=4$$\to$$F_{\rm e}=3$ in the D$_1$ line. Unfortunately, the \textit{lin}$\perp$\textit{lin} configuration is sensitive to a spectral resolution of the excited-state hfs levels. Therefore, it cannot be successfully employed for miniaturized vapor cells where narrow light beams and a relatively high buffer gas pressure is required. Besides, the configuration exhibits a shift effect \cite{EIAShift}, which can affect deeply the magnetic field measurements.

\begin{figure}[!t]
\centering
\includegraphics[width=\linewidth]{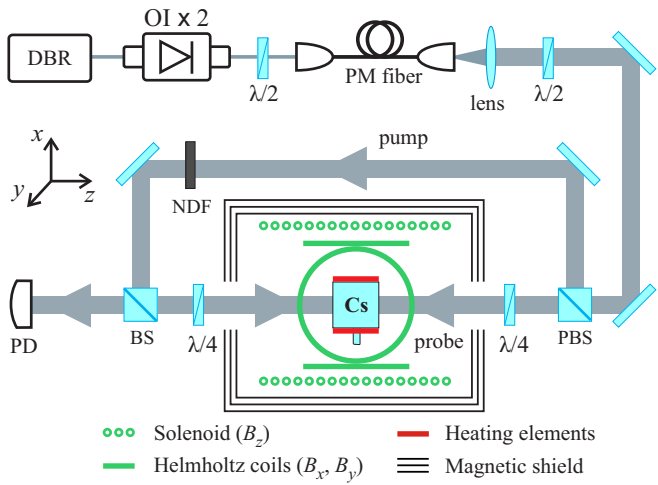}
\caption{Experimental setup (see details in the text).}
\label{fig:setup}
\end{figure}

Here we examine a configuration composed of counterpropagating pump and probe beams with opposite circular polarizations ($\sigma^+\sigma^-$). It should be noted that similar configurations have already brought relevant results for developing atomic clocks \cite{Yudin2017,EIAClock}. The same light field geometry was studied in \cite{Gozzini2017} in a cm-size K vapor cell with antirelaxation coating of the walls. In the present study we show that a buffer-gas-filled Cs vapor cell can provide much better contrast-to-width ratio (CWR) of the GSHE resonance due to the spectral resolution of ground-state hfs levels. The considered configuration keeps all advantages of the \textit{lin}$\perp$\textit{lin} scheme, while it allows using a higher buffer gas pressure and does not exhibit the shift effect \cite{EIAShift}. Also, we compare the obtained results with those registered in a standard Hanle scheme using a single light wave in the same vapor cell to show the benefits from using the $\sigma^+\sigma^-$ configuration.

\section{\label{sec:Setup}Experimental Setup}

The experimental setup is shown in fig. \ref{fig:setup}. The output beam of a distributed Bragg reflector (DBR) diode laser with a radiation wavelength of $\lambda$$=$$\,894.6$~nm (Cs D$_1$ line) is passed through two Faraday optical isolators (OI). Subsequently, it is sent to a polarization maintaining optical fiber. A half-wave plate ($\lambda$/2) before the fiber is used to adjust the linear polarization of the beam. A lens is placed after the fiber to collimate the beam. The combination of a half-wave plate placed after the fiber and a polarizing beam splitter (PBS) is used to redistribute the total light power between the pump and probe beams, which have orthogonal linear polarizations after the PBS. A beam splitter (BS) is used to direct the pump beam into the cell without affecting its polarization. The beam diameters in the cell ($1/e^2$) are $\approx\,3$~mm. A set of neutral density filters (NDF) is used to tune the pump-wave power. Two quarter-wave plates ($\lambda$/4) are placed on both sides of the cell to produce the $\sigma^+\sigma^-$ configuration.

\begin{figure}[!b]
\centering
\includegraphics[width=0.9\linewidth]{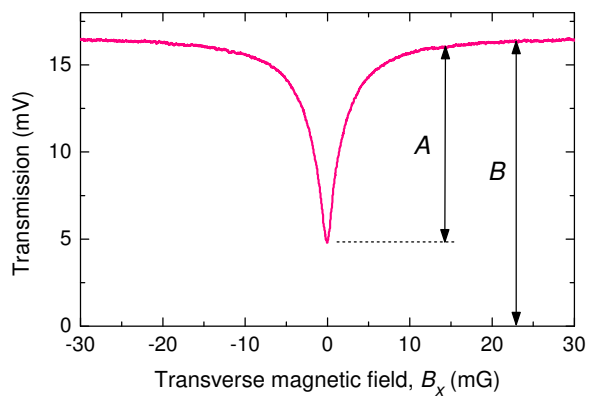}
\caption{Level-crossing EIA resonance in the probe-wave transmission. $P_c$$\approx\,$$800$~$\mu$W, $P_p$$\approx\,$$5$~$\mu$W, $T_{\rm cell}$$\approx\,$$55^\circ$C.}
\label{fig:EIA}
\end{figure}

The cesium vapor cell is filled with a neon buffer gas ($\approx\,120$~Torr) and it has a cubic shape with inner volume of $5\times5\times5$$\,=\,$$125$~mm$^3$. The cell is heated by an ac electric current ($100$~kHz) applied to special heating elements. The heating process does not have a visible effect on the LC resonances. A three-layer $\mu$-metal magnetic shield is utilized to reduce the stray field to $\approx\,0.1$~mG in the cell area. The Doppler absorption profiles corresponding to the transitions $F_g$$=$$4$$\to$$F_e$$=$$3$ and $F_g$$=$$4$$\to$$F_e$$=$$4$ are significantly broadened due to buffer-gas collisions and, therefore, they are overlapped. And the laser frequency is tuned manually to the center of the combined absorption profile. We use a solenoid to produce a longitudinal magnetic field (${\bf B}_z$$||$${\bf k}$) and two pairs of Helmholtz coils to produce transverse components (${\bf B}_x$, ${\bf B}_y$).

\section{\label{sec:Measur}Measurements and Discussions}

\begin{figure}[!t]
\centering
\includegraphics[width=0.9\linewidth]{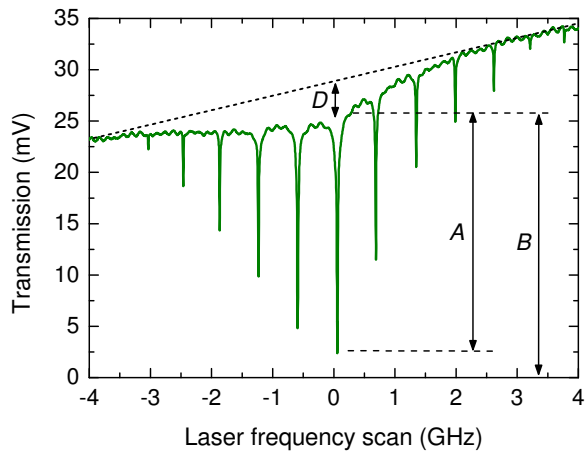}
\caption{Probe-wave transmission versus the laser frequency scan around the transitions $F_g$$=$$4$$\to$$F_e$$=$$3,\,4$ with simultaneous scan of the transverse magnetic field around zero. The frequencies of the scans are: $f_{\rm mag}$$=$$15$~Hz, $f_{\rm las}$$=$$0.5$~Hz. $P_c$$\approx\,$$5.13$~mW, $P_p$$\approx\,$$5$~$\mu$W, $T_{\rm cell}$$=$$60^\circ$C. The dotted line represents the signal level without absorption in the cell. The central spike contrast is $\approx\,92\%$.}
\label{fig:HanleDoppler}
\end{figure}

An example of the GSHE resonance that can be observed in our scheme is shown in fig.\ref{fig:EIA}. The physical grounds for observing the high-contrast EIA-type LC resonances are similar to those discussed in details in \cite{JPhysB2019} for the configuration with linearly polarized waves. Here we will just briefly describe them. Let us consider a resonant interaction of the $\sigma^+$ polarized pump wave with the optical transitions $F_g$$=$$4$$\to$$F_e$$=$$3,\,4$. We assume the quantization axis $z$ to be directed along the wave vectors. At $B_x$$=\,$$0$, a significant part of the atoms are accumulated on the edge Zeeman sub-level $|F_g$$=$$4,\,$$m_g$$=$$4$$\rangle$ where they do not interact with the pump field. It can be referred to as a trap state. The existence of such state leads to a small pump-wave absorption in the cell. The counterpropagating $\sigma^-$ probe wave, on the contrary, experiences a strong interaction with the trap state that means a high probe-wave absorption (see fig.\ref{fig:EIA}). In the figure, $A$ denotes the resonance height, while $B$ stands for the magnitude of the background transmission. $P_c$ and $P_p$ denote the pump (control) and the probe wave powers, respectively. Then, if $B_x$$\ne\,$$0$, the Zeeman sub-levels are mixed by the magnetic field, leading to the destruction of the trap state and thus to an increase in the pump-wave absorption. The latter, in turn, results in transferring most atoms to the non-resonant ground-state level $F_g$$=$$3$ via the optical pumping process. Being on this level, the atoms do not scatter both the pump and the probe light waves. This is how the EIA resonance is created. Obviously, increasing the buffer gas pressure implies increase in the time of light-atom interaction. It leads, on the one hand, to a decrease in the resonance linewidth and, on the other hand, to an increase in the efficiency of all the optical pumping processes mentioned above. The described mechanism yields both high-contrast and narrow-linewidth LC resonances, hence the necessity of spectral resolution of the ground-state levels.

\begin{figure}[!b]
\centering
\includegraphics[width=0.88\linewidth]{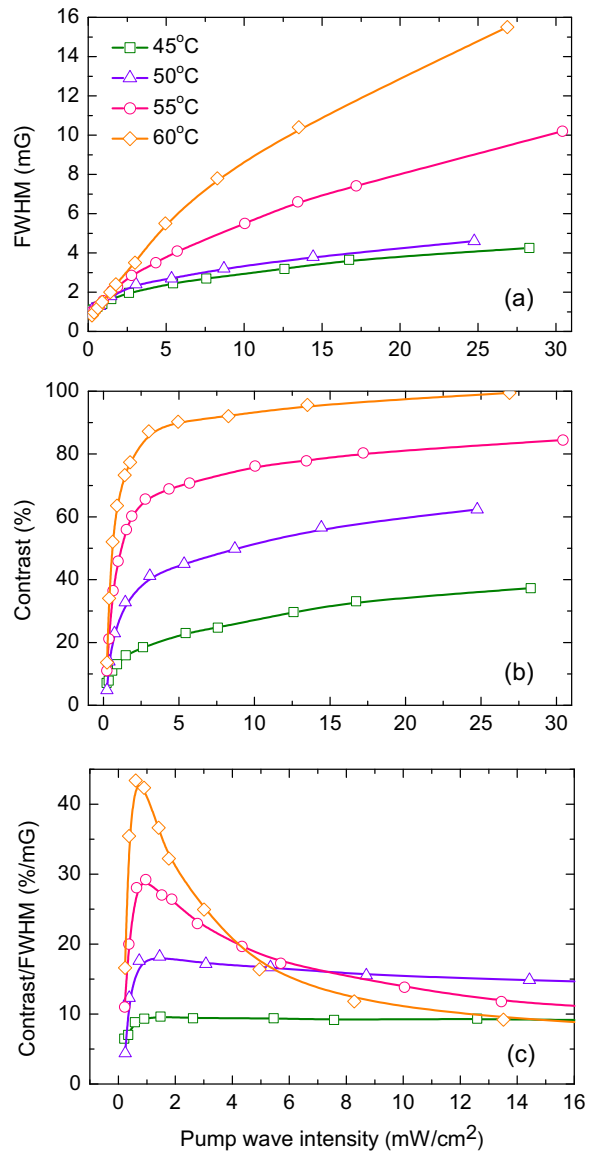}
\caption{Properties of the level-crossing EIA resonance versus the pump-wave intensity: (a) linewidth, (b) contrast, and (c) contrast-to-width ratio. The results are obtained for different cell temperatures. $P_p$$\approx\,$$5$~$\mu$W.}
\label{fig:Parameters}
\end{figure}

The plot in fig.\ref{fig:HanleDoppler} demonstrates the probe-wave transmission versus the laser frequency scan around the transitions $F_g$$=$$4$$\to$$F_e$$=$$3,\,4$, while simultaneously scanning $B_x$ around zero. As seen, the EIA LC resonances can have a very high contrast with respect to both the Doppler absorption profile ($D$) and the background transmission ($B$). This curve can be compared, for instance, with that presented in fig.2 in Ref.\cite{Gozzini2017} for potassium vapors. The comparison reveals a much higher contrast and CWR in the case of Cs vapors. In the current work, we focus on measuring the full width at half maximum (FWHM) of the resonance, its contrast defined as $C$$=$$(A/B)\times100$\% (see fig.\ref{fig:EIA}), and the contrast-to-width ratio ($C/$FWHM). The latter is especially important for applications in magnetometry.

\begin{figure}[!t]
\centering
\includegraphics[width=0.95\linewidth]{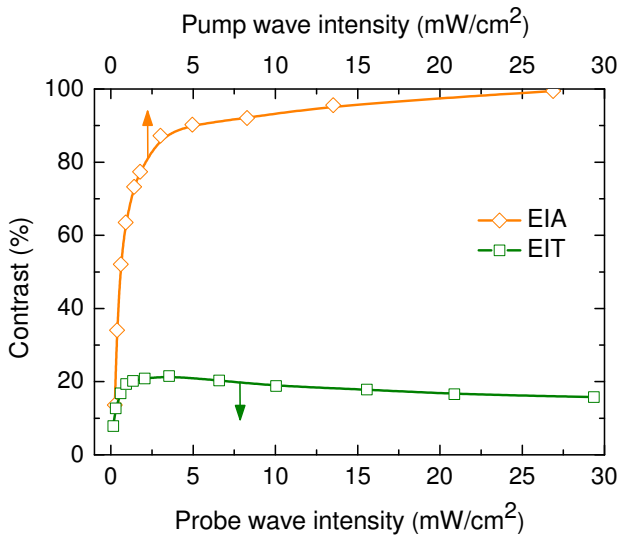}
\caption{Comparison of the contrast of the EIA resonance in the $\sigma^+\sigma^-$ configuration ($P_p$$\approx\,$$5$~$\mu$W) with the contrast of the EIT resonance in the standard Hanle configuration. $T_{\rm cell}$$=$$60^\circ$C.}
\label{fig:EIAvsEIT}
\end{figure}

Fig.\ref{fig:Parameters} reveals the influence of the pump-wave intensity ($I_c$$=$$4P_c/\pi d^2$) on the resonance parameters. The power broadening of the resonance is shown in fig.\ref{fig:Parameters}a. The zero power limit gives FWHM$\approx0.6$~mG ($60$~nT). The linewidth dependence does not show a significant difference between the regimes with $T_{\rm cell}$$=\,$$45^\circ$C and $50^\circ$C. However, further increase in the temperature results in additional broadening, because the medium acquires large optical density. The temperature increase, on the contrary, helps to observe an extremely high contrast of the resonance, reaching $\approx\,99\%$ (see fig.\ref{fig:Parameters}b). The competition between the linewidth and the contrast dependencies leads to the formation of an extremum in the contrast-to-width ratio (fig.\ref{fig:Parameters}c). This extremum shows that the optimal pump-wave intensity is not high ($\approx\,0.7$~mW/cm$^2$). We should note that the observed CWR$\approx45$~\%/mG is relatively high, especially taking into account the cell size. For instance, in \cite{Gozzini2009}, the authors measured a CWR of the EIT resonance of $\approx65/13=5$~\%/mG in a 5 cm long K buffer-gas cell. In \cite{Kim2011}, the authors reported observation of an EIA LC resonance as narrow as $0.55$~mG in a 5 cm long Rb vapor cell with antirelaxation coating. In spite of the small linewidth, a contrast of only $0.5$\% was achieved, yielding CWR$\approx0.9$~\%/mG.

\begin{figure}[!t]
\centering
\includegraphics[width=0.95\linewidth]{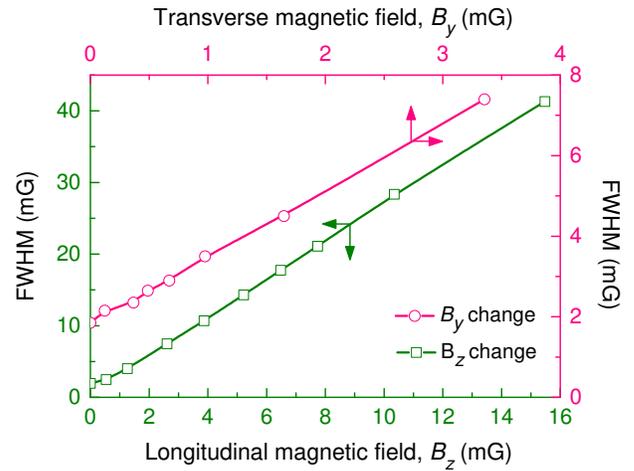}
\caption{EIA resonance linewidth as a function of the $B_y$ and $B_z$ components of an ambient magnetic field. $P_c$$\approx\,$$265$~$\mu$W, $P_p$$\approx\,$$5$~$\mu$W, $T_{\rm cell}$$=$$50^\circ$C.}
\label{fig:BzBx}
\end{figure}

Fig.\ref{fig:EIAvsEIT} reveals the benefits of using the $\sigma^+\sigma^-$ light-field configuration rather than the standard Hanle scheme where only one circularly polarized wave is used. In the latter case, an EIT LC resonance is observed. To compare the two configurations, we used the same setup as in fig.\ref{fig:setup}, but with the pump wave switched off. As seen from fig.\ref{fig:EIAvsEIT}, the EIA resonance contrast demonstrates a fivefold increase with respect to the EIT one. For applications in magnetometry, it is important that the linewidth of a magneto-optical resonance demonstrates a linear sensitivity to variations in the magnetic field. Obviously, a constant component $B_x^\prime$ of an ambient field will lead to just a shift of the resonance that can be measured. Fig.\ref{fig:BzBx} shows that the other two components provide a linear broadening of the resonance.

\section{\label{sec:Concl}Conclusions}

We reported a substantial improvement of the standard Hanle configuration for observing level-crossing resonances in buffered vapor cells. The proposed technique yields ultrahigh-quality resonances in a small low-temperature vapor cell using only two transparent windows. Therefore, the results can be applied for developing a miniaturized low-power magnetic field sensor with enhanced sensitivity.

\begin{acknowledgments}

We thank Russian Science Foundation (17-72-20089). The work, in part, has also been supported by Russian Foundation for Basic Research (20-52-18004) and Bulgarian National Science Fund (KP-06-Russia/11) in the framework of a joint research project. I.S. Mesenzova thanks Ministry of Science and Higher Education of the Russian Federation (Presidential Scholarship SP-269.2021.3).

\end{acknowledgments}

\end{document}